\documentclass[a4paper,11pt]{article}
\ifdefined\pdfoutput\pdfoutput=1\fi

\usepackage{jheppub}
\makeatletter\gdef\@fpheader{}\makeatother
\usepackage{amsfonts,bm}
\usepackage{physics}
\usepackage{xcolor}
\usepackage{indentfirst}
\usepackage{orcidlink}
\usepackage{subfigure}

\hypersetup{pdftitle={Landau-Ginzburg Theory of a Holographic Superconductor}, pdfauthor={Sejin Kim, Yunseok Seo}}
\title{Landau–Ginzburg Theory of a Holographic Superconductor: A Horizon Perspective on the Zero-Temperature Ground State
}

\author[a]{Sejin Kim\,\orcidlink{0000-0003-1176-4199}}
\author[b]{Yunseok Seo\,\orcidlink{0000-0001-5652-5736}}

\affiliation[a]{Center for Artificial Intelligence and Natural Sciences, Korea Institute for Advanced Study, Seoul 02455, Korea}
\affiliation[b]{College of General Education, Kookmin University, Seoul 02707, Korea}

\emailAdd{sejin@kias.re.kr}
\emailAdd{yseo@kookmin.ac.kr}

\abstract{%
We investigate the effect of a quartic scalar self-interaction, $\lambda|\phi|^4$, on a holographic superconductor in which non-linear electrodynamic couplings generate a superconducting dome containing an extended Lifshitz regime.
The quartic interaction leaves the linearized scalar dynamics around the normal phase unchanged while modifying the non-linear solutions where the scalar field is large.
As $\lambda$ increases, the solutions within the dome pass through three stages. The Lifshitz region shrinks until its scaling solution disappears at $\lambda^*$, leaving a no-AdS region, beyond whose edge the flows end before reaching any asymptotic region, and above $\lambda_c$ the dome consists entirely of asymptotically AdS hairy black holes.
At the bottom of the dome we identify the point where the horizon value of the scalar field is largest, which we refer to as the critical point.
There this value lies at the deepest symmetry-breaking minimum that the horizon effective potential, of Landau–Ginzburg form, can have.
Unlike a conventional Landau–Ginzburg description, this interpretation rests on the scalar field at the horizon rather than on the boundary order parameter.
Below $\lambda^*$, the constant scalar of the Lifshitz fixed point sits instead at the minimum of the effective potential of its asymptotic scaling region.
}

\begin{document}
\maketitle

\section{Introduction}
Phase transitions in condensed matter systems are most naturally described in terms of an order parameter that distinguishes the symmetric and broken-symmetry phases.
This picture was first formalized by Landau, who showed that the free energy of a system near a continuous phase transition can be expanded in powers of an order parameter, with the transition itself driven by a symmetry-breaking mechanism~\cite{Landau1937}.
One of the most celebrated realizations of this idea is the Ginzburg--Landau theory of superconductivity, in which a complex order parameter describes the condensation of Cooper pairs below the critical temperature~\cite{GinzburgLandau1950,BCS1957}.

The holographic superconductor was constructed as a gravitational dual of precisely this Landau--Ginzburg mechanism.
In the model of Gubser, Hartnoll, Herzog and Horowitz, a charged complex scalar field in an asymptotically AdS black hole background becomes unstable and condenses below a critical temperature, and the resulting scalar profile is dual to the order parameter of a strongly coupled boundary superconductor~\cite{Gubser:2008px,Hartnoll:2008vx,Hartnoll:2008kx,Cai:2015cya,Horowitz:2008bn,Hartnoll:2009sz,Herzog:2009xv,Hartnoll:2016apf}.

The stability of a scalar field in an AdS background is governed by the Breitenlohner--Freedman (BF) bound, which allows the effective mass-squared of the scalar to be negative, down to a finite negative value, without rendering the background unstable~\cite{Breitenlohner:1982jf}.
In the holographic superconductor, the mechanism behind the instability can be understood precisely in terms of this bound.
Although the scalar field satisfies the BF bound with respect to the full asymptotically AdS geometry, its charge coupling to the background electric field provides an additional negative contribution to its effective mass, and at sufficiently low temperature this contribution becomes large enough that the effective mass violates the BF bound associated with the near-horizon AdS$_2\times\mathbb{R}^2$ throat geometry of the extremal black hole~\cite{Cai:2015cya,Faulkner:2009wj}.
This violation of the near-horizon BF bound signals a genuine instability of the scalar field and triggers its condensation, so that the endpoint of the instability is a new, hairy black hole solution dual to the superconducting phase.

Another direction in which the holographic superconductor has been made more realistic is the introduction of momentum relaxation.
In the original construction, translational invariance of the boundary theory is preserved, so the momentum of the charge carriers is an exactly conserved quantity.
As a consequence, the real part of the AC conductivity contains a delta function at zero frequency and the DC conductivity is formally infinite even in the normal, non-superconducting phase.

Real materials, by contrast, always contain impurities or a lattice that breaks translational symmetry, allowing momentum to relax and giving rise to a finite, generically temperature-dependent DC resistivity.
To capture this effect holographically, translational symmetry is broken explicitly in the bulk. One way is a spatially dependent source, such as a periodic lattice deformation of the metric or the gauge field~\cite{Horowitz:2012ky}. A more tractable one turns on bulk massless scalar fields (axions) linear in the boundary spatial coordinates, which break translations while preserving homogeneity~\cite{Andrade:2013gsa,Baggioli:2021xuv}. A third treats the bulk graviton as massive~\cite{Vegh:2013sk,Davison:2013jba,Blake:2013owa}.

This axion-based momentum-relaxation framework has subsequently been applied to a wide range of strongly coupled transport phenomena, including coherent/incoherent metallic transport, spectral functions, and transitions from metal to insulator driven by an interplay between momentum relaxation and impurity or magnetic-field effects~\cite{Kim:2014bza,Jeong:2019lss,Seo:2023dwr,Kim:2015wba,Seo:2016vks,Song:2019rnf,Kim:2019lxb,Kim:2020ozm}.
Holographic models of the same kind have also been applied to strong correlation on the surfaces of topological insulators, to the Mott transition and to flat band and Fermi arc states~\cite{Seo:2017oyh,Seo:2018hrc,Oh:2020cym}.

More directly relevant to the present context, the framework has also been incorporated into the holographic superconductor itself to compute momentum-relaxation-dependent optical and DC conductivities and to test holographic analogues of Homes' law~\cite{Ling:2014laa,Kim:2015dna,Kim:2016hom}.
Incorporating such momentum-relaxation mechanisms into the holographic superconductor removes the unphysical infinite DC conductivity of the normal phase and allows one to compute finite, momentum-relaxation-dependent transport coefficients across the superconducting transition, bringing the holographic model into closer contact with the transport properties measured in real superconducting and strange-metal materials.

While the holographic superconductor elegantly reproduces the qualitative features of a symmetry-breaking phase transition on the boundary, such as the divergence of the DC conductivity and the opening of a gap, the resulting boundary phase diagram is much simpler than what is observed experimentally.
In particular, the original holographic superconductor develops a condensate at arbitrarily low temperature for any nonzero chemical potential, so the superconducting phase always extends down to $T=0$.
Real superconducting materials, by contrast, exhibit a much richer variety of phase diagrams as a function of an external tuning parameter such as doping or pressure.
Many unconventional and high-$T_c$ materials, most famously the cuprates, show a characteristic dome-shaped superconducting region bounded on both sides in the doping-temperature plane, together with competing normal, pseudogap and strange-metal phases~\cite{Keimer:2015nature,Fradkin:2015RMP}.

To bring the holographic model closer to this phenomenology, non-linear electrodynamic (NED) interactions between the complex scalar field and the bulk $U(1)$ gauge field were introduced~\cite{Seo:2023kyp}.
Such an interaction modifies the effective mass of the scalar field through its coupling to the gauge field strength, and it is precisely this field-strength-dependent effective mass that controls the onset of the instability and hence the shape of the phase diagram.
By tuning the coefficients of the non-linear interaction term, the condensation can be made to occur only within a finite window of the tuning parameter, reproducing a dome-shaped superconducting region qualitatively similar to that seen in real materials~\cite{Seo:2023kyp,Kim:2024ndba}.

Earlier routes to richer holographic phase diagrams used general couplings of the scalar~\cite{Franco:2009yz}.
Competing condensed phases gave another route~\cite{Horowitz:2010jq}.
Superconducting domes have been obtained from doping and from broken translations~\cite{Ganguli:2013oya,Baggioli:2015dwa,Kiritsis:2015hoa,Cai:2020nyd}.
Non-linear electrodynamics of the Born--Infeld type has also been coupled to the holographic superconductor~\cite{Jing:2010zp,Zhao:2012cn}.

A further characteristic feature of the model of Ref.~\cite{Seo:2023kyp} is the emergence of a ``quantum critical region'' inside the superconducting dome.
In a portion of the condensed phase, horizon data give flows that are no longer asymptotically AdS but instead approach a Lifshitz-scaling geometry far from the horizon, signaling an emergent, anisotropic scaling symmetry reminiscent of quantum critical behavior~\cite{Seo:2023kyp,Kachru:2008yh,Taylor:2008tg,Gubser:2009cg,Horowitz:2009ij,Goldstein:2009cv,Charmousis:2010zz,Iqbal:2011aj,Horowitz:2011gh,Ge:2019fnj,Song:2020ojc}.
Interestingly, this Lifshitz region is found to occupy a substantial fraction of the superconducting dome, suggesting that quantum-critical physics may be intimately tied to the dome-shaped phase structure itself, much as it is conjectured to underlie the strange-metal behavior observed near the optimal doping of cuprate superconductors~\cite{Keimer:2015nature,Fradkin:2015RMP}.

In this work, we extend this non-linear-electrodynamics holographic superconductor by including an additional $\lambda|\varphi|^4$ self-interaction term for the complex scalar field, a term that has been considered in holographic superconductors without the non-linear interaction~\cite{Franco:2009yz,Gubser:2009cg,Zhao:2022jvs}.
For a potential of this W shape, the zero-temperature limit of the holographic superconductor has an infrared region in which the scalar sits at the symmetry-breaking minimum and conformal symmetry re-emerges or, when the charge of the scalar is small, an infrared with Lifshitz-like scaling~\cite{Gubser:2009cg}.
In the model of Ref.~\cite{Seo:2023kyp} the electric field enters the quadratic coefficient of the potential that the scalar sees, so the depth of this minimum depends on the field at the horizon, and this dependence is what singles out the critical point found below.
Since the scalar field vanishes at the asymptotic AdS boundary, this quartic term does not modify the leading near-boundary expansion and therefore leaves the scaling dimensions and the definitions of the source and the vacuum expectation value of the order parameter unchanged.
However, because the scalar field is generically nonzero throughout the bulk and grows large near the horizon, the $\lambda|\varphi|^4$ term can substantially affect the near-horizon (IR) geometry, and hence the structure of the hairy black hole solutions, the extent of the quantum critical (Lifshitz) region, and the resulting shape of the phase diagram.

We find that the quartic potential removes the Lifshitz region in three stages.
Below a coupling $\lambda^*$ the region shrinks, its upper edge moving to lower temperature.
At $\lambda^*$ the Lifshitz fixed point merges with a second, repelling fixed point and ceases to exist.
Between $\lambda^*$ and a larger coupling $\lambda_c$ the region left behind in the dome keeps shrinking, but horizon data beyond its inner edge, which lie beyond the wall at which the response of the non-linear electrodynamics saturates, give flows that close off before reaching any asymptotic region, and we call it the no-AdS region.
For $\lambda > \lambda_c$ this region has closed, no source free solution reaches the wall any more, and the whole dome is filled with source free hairy black holes with an AdS boundary.
Both couplings depend on $m^2$, $q$, $\gamma_2$ and $\gamma_4$, and their values for the parameters of this work are given in (\ref{eq:lstarvalues}) and (\ref{eq:lamc}).

In the filled dome the scalar at the horizon sees a potential of Landau--Ginzburg form, $m_{eff}^2 \phi^2 + \lambda \phi^4$, whose mass term is set by the electric field at the horizon, and at the critical point, the point at the bottom of the dome where the horizon value of the scalar is largest, that value sits at the deepest minimum this double well can have.
Below $\lambda^*$, the constant scalar of the Lifshitz fixed point sits at the minimum of a well of the same form, as the scalar equation requires of a frozen scalar, with the charge of the scalar now in the mass term.

Section \ref{sec:model} sets up the model with the quartic term and classifies its solutions.
Section \ref{sec:stages} starts from the phase diagram at $\lambda = 0$, identifies the wall, the electric field at which the response of the non-linear electrodynamics saturates and the Maxwell equation degenerates, and follows the interior of the dome through the three stages.
Section \ref{sec:veff} analyzes the effective potential of the scalar at the horizon and in the scaling region, and relates the critical point and the Lifshitz fixed point to their minima.
Section \ref{sec:discussion} discusses what is left open, Appendix \ref{sec:lambdamax} determines $\lambda^*$, and Appendix \ref{sec:beyondwall} follows the flows beyond the wall.

\section{\texorpdfstring{Landau--Ginzburg theory for a holographic superconductor}{Landau-Ginzburg theory for a holographic superconductor}}\label{sec:model}

\subsection{Action and equations of motion}\label{sec:action}

We start from the holographic superconductor of Ref.~\cite{Seo:2023kyp}, in which a non-linear interaction couples the complex scalar field to the $U(1)$ gauge field strength, and add to it a quartic self-interaction of the scalar, the Landau--Ginzburg (LG) extension of the model.
The bulk action is given by
\begin{align}\label{eq:Bulk_Action}
S_{Bulk} = \frac{1}{16\pi G} \int d^4 x\sqrt{-g}\left( R+ \frac{6}{L^2} - \frac{1}{2}\sum_{i=1}^2\left(\partial\chi^{i}\right)^2 - \frac{1}{4}F^2 -|D\varphi|^2 -V(\varphi, F^2) \right),
\end{align}
where the scalar potential collects the mass term, the quartic self-interaction and the non-linear electrodynamic interaction,
\begin{align}\label{eq:Veffcov}
V(\varphi, F^2) = m^2 |\varphi|^2 + \lambda |\varphi|^4 +\frac{1}{4}  |\varphi|^2\left(\gamma_2 F^2 +\gamma_4  L^2  (F^2)^2\right).
\end{align}
The last term is the non-linear electrodynamic interaction of Ref.~\cite{Seo:2023kyp}, and it enters only as a field-dependent mass of the scalar.

The linear axion fields $\chi^i$ realize the effect of impurities, which relax momentum, and the complex scalar $\varphi$ the superconducting order.
The electromagnetic field is given by $F=dA$, and $F^2$ denotes $F_{MN}F^{MN}$.
The scalar carries the charge $q$, and $D_M = \partial_M - i q A_M$ is its gauge covariant derivative.
For $\lambda = 0$ the action reduces to that of Ref.~\cite{Seo:2023kyp}, and we refer the reader there for the details of the calculation.

From the bulk action (\ref{eq:Bulk_Action}), we obtain the Einstein, Maxwell, scalar and axion equations,
\begin{align}
&R_{M N} -\frac{1}{2}g_{M N} {\cal L} -\frac{1}{2}\sum_{i=1}^{2} \partial_M \chi^i \partial_N \chi^i -\frac{1}{2}\left(1+\gamma_2 |\varphi|^2+2 \gamma_4 L^2 |\varphi|^2 F^2\right)F_M{}^P F_N{}_P \nonumber\\
&\hphantom{R_{M N}} - \frac{1}{2}\left[D_M\varphi (D_N\varphi)^*+D_N\varphi (D_M\varphi)^*\right]=0 , \label{eq:Einstein}\\
&\nabla_M \left[ \left( 1 + \gamma_2 |\varphi|^2 + 2 \gamma_4 L^2 |\varphi|^2 F^2\right) F^{MN} \right] = i q \left(\varphi^* D^N \varphi - \varphi (D^N \varphi)^* \right) , \label{eq:Maxwell}\\
&D^2 \varphi -\left(m^2+\frac{1}{4}\gamma_2 F^2 +\frac{1}{4} \gamma_4 L^2 (F^2)^2 \right) \varphi -2 \lambda |\varphi|^2 \varphi = 0 , \label{eq:scalar}\\
&\nabla^2 \chi^i = 0 , \label{eq:axion}
\end{align}
where ${\cal L}$ is the Lagrangian density of (\ref{eq:Bulk_Action}), through which the $\lambda$ term enters the Einstein equation.
The $\lambda$ term does not enter the Maxwell equation.

We take the usual black-brane ansatz,
\begin{align}\label{Ansatz}
&ds^2 = - U(r) e^{2 W(r)} dt^2 + \frac{r^2}{L^2} \left(dx^2 + dy^2\right) + \frac{dr^2}{U(r)}\nonumber\\
&\chi^1 = \kappa\, x, ~\chi^2 = \kappa\, y~,~A= A_t(r) dt~,~\varphi = \phi(r) e^{i \tilde{\phi}(r)}\,,
\end{align} 
where the axion charge $\kappa$ sets the strength of momentum relaxation, and $\phi$ and $\tilde{\phi}$ are the modulus and the phase of $\varphi$.
With this ansatz the axion equation is automatically satisfied.
The $r$-component of the Maxwell equation requires the phase $\tilde{\phi}(r)$ in (\ref{Ansatz}) to be constant, and we set $\tilde{\phi}(r) = 0$.

Substituting the ansatz (\ref{Ansatz}) into (\ref{eq:Einstein}), (\ref{eq:Maxwell}) and (\ref{eq:scalar}) gives six equations for the component fields, of which four are independent.
We scale all fields and $r$ with $L$ and $r_h$ such that the equations of motion are dimensionless, and write them with the local electric field strength and the displacement,
\begin{align}\label{eq:E}
E \equiv e^{-W} A_t' , ~~~~ D \equiv \left( 1 + \gamma_2 \phi^2 \right) \left( 1 - \frac{E^2}{3 E_{\rm sat}^2} \right) E , ~~~~ E_{\rm sat}^2 \equiv \frac{1 + \gamma_2 \phi^2}{12 \gamma_4 \phi^2} .
\end{align}
On the ansatz $F^2 = - 2 E^2$, and $D = \partial {\cal L} / \partial E$ is the derivative of the Lagrangian density of (\ref{eq:Bulk_Action}) with respect to the field.
The displacement grows with $E$ only up to $E = E_{\rm sat}$, where $\partial_E D = 0$ and it saturates, hence the subscript.
The equations are
\begin{align}
&\phi'' + \left( \frac{2}{r} + W' + \frac{U'}{U} \right) \phi' - \frac{\phi}{U} \left( m^2 - \frac{\gamma_2}{2} E^2 + \gamma_4 E^4 - \frac{q^2 e^{-2 W} A_t^2}{U} + 2 \lambda \phi^2 \right) = 0 , \label{eom1}\\
&\left( r^2 D \right)' = \frac{2 q^2 r^2 \phi^2 e^{-W} A_t}{U} , \label{eom2}\\
&U' = 3 r - \frac{\kappa^2}{2 r} - \frac{U}{2 r} \left( 2 + r^2 \phi'^2 \right) - \frac{q^2 r \phi^2 e^{-2 W} A_t^2}{2 U} \nonumber\\
&\hphantom{U' =} - \frac{r}{4} \left[ \left( 1 + \gamma_2 \phi^2 \right) E^2 - 6 \gamma_4 \phi^2 E^4 + 2 m^2 \phi^2 + 2 \lambda \phi^4 \right] , \label{eom3}\\
&W' = \frac{r}{2} \phi'^2 + \frac{q^2 r \phi^2 e^{-2 W} A_t^2}{2 U^2} . \label{eom4}
\end{align}
On the ansatz the non-linear electrodynamic part of (\ref{eq:Veffcov}) is $|\varphi|^2 \left( - \gamma_2 E^2 / 2 + \gamma_4 E^4 \right)$.

The equations of motion are invariant under $W \rightarrow W + c$, $A_t \rightarrow e^{c} A_t$, which together with $t \rightarrow e^{-c} t$ leaves the metric (\ref{Ansatz}) and the field strength $E$ unchanged.
For the solutions with an asymptotically AdS region we fix $c$ by $W(\infty) = 0$, so that $t$ is the time of the boundary theory.

At the horizon, the gauge field $A_t (r)$ and the metric function $U(r)$ should vanish, while the scalar field $\phi(r)$ and $W(r)$ are finite.
From the regularity condition of the equations of motion at the horizon, we get the horizon conditions for each field as
\begin{align}\label{eq:horizon}
&A_t (r_h) =0,~~~~~~~E (r_h) = E_h \cr
&\phi (r_h) = \phi_h,~~~~~~\phi'(r_h) = \frac{\phi_h (2 m^2 - \gamma_2 \, E_h^2 +2 \gamma_4  \,E_h^4 +4 \lambda \phi_h^2 )}{2 U'(r_h)} \cr
&U(r_h) = 0,~~~~~~ U'(r_h) = \frac{1}{4}\Big( 12 - E_h^2 -2 m^2 \phi_h^2  - 2\kappa^2 \cr
&\hphantom{U(r_h) = 0,~~~~~~ U'(r_h) = \frac{1}{4}\Big(} - \gamma_2 \, \phi_h^2 \, E_h^2 + 6 \gamma_4 \, \phi_h^2 \, E_h^4 -2 \lambda \phi_h^4 \Big), \cr
&W(r_h) = 0,~~~~~~W'(r_h) = \frac{1}{2} \phi'(r_h)^2 + \frac{q^2 \phi_h^2 E_h^2}{2 U'(r_h)^2}.
\end{align}

In the presence of the black hole horizon, the temperature and the entropy density of the boundary system are defined as
\begin{align}\label{eq:TS}
T= \frac{1}{4\pi} U'(r_h) e^{W (r_h)},~{\cal S} =\frac{r_h^2}{4GL^2}\,.
\end{align}
At fixed horizon data the quartic coupling lowers $U'(r_h)$ through the last term of its expression in (\ref{eq:horizon}), and with it the temperature (\ref{eq:TS}).

\subsection{Three types of solutions}\label{sec:solutions}

Depending on the initial conditions imposed at the black hole horizon, the background solutions of this model can be classified into three types, namely the RN black hole, the hairy black hole with an asymptotically AdS geometry, and the hairy black hole with an asymptotically Lifshitz geometry.
 
The first type, and the simplest solution of the model, is the one with a vanishing scalar field.
In this case all the non-linear electrodynamic interaction terms and the quartic term vanish identically, so that the action reduces to that of the usual Einstein--Maxwell system with the axions, and the solution is the Reissner--Nordstr\"{o}m (RN) black hole with momentum relaxation,\footnote{We set $16\pi G=1$ and $L=1$ for all the following calculations.}
\begin{align}
U(r)&=r^2-\frac{\kappa^2 }{2}-\frac{M }{r}+\frac{{\cal Q}^2 }{4 r^2}\,, \cr
A_t(r)&=\mu  -\frac{{\cal Q} }{r}, \cr
\phi(r)&=W(r)=0\,.
\end{align}
The mass parameter is fixed by the condition $U(r_h)=0$ at the horizon,
\begin{align}
M= r_h^3+\frac{ {\cal Q}^2}{4 r_h}-\frac{\kappa ^2 r_h}{2 }\,,
\end{align}
where $r_h$ denotes the location of the horizon.
The Hawking temperature and the entropy density are then given by
\begin{align}
T = \frac{1}{4 \pi  } \left(3 r_h-\frac{ {\cal Q}^2}{4 r_h^3}-\frac{\kappa ^2}{2 r_h}\right)~,~{\cal S} = 4 \pi r_h^2\,.
\end{align}
This solution corresponds to the normal, metallic phase of the system, whose transport properties are studied in Refs.~\cite{Andrade:2013gsa,Kim:2014bza}.
 
The second type of solution is the hairy black hole with asymptotically AdS geometry,
\begin{align}
U(r) \Big|_{r \rightarrow \infty} \sim r^2 +\cdots, ~~~~~W(r) \Big|_{r \rightarrow \infty} \sim 0 + \cdots.
\end{align}
Because of the non-linear electrodynamic interaction, the electric flux in the bulk is carried by the displacement $D$ of (\ref{eq:E}) rather than by $E$, but $D \rightarrow E$ where the scalar vanishes.
As usual, the charge carrier density and the conjugate chemical potential are read off from the asymptotic expansion of the gauge field,
\begin{align}\label{asymAt}
A_t (r) \Big|_{r \rightarrow \infty} \sim \mu - \frac{{\cal Q}}{r} +\cdots,
\end{align}
where the physical charge carrier density is given by ${\cal Q} = \lim_{r \rightarrow \infty} r^2 A_t'(r)$.

With these asymptotic behaviors of the metric and the gauge field, the scalar field admits the near-boundary expansion
\begin{align}\label{aymphi}
\phi (r) \Big|_{r \rightarrow \infty} \sim \frac{{\cal O}_{\phi}^{-}}{r^{\Delta_{-}}} + \frac{{\cal O}_{\phi}^{+}}{r^{\Delta_{+}}} +\cdots, ~~~~~\Delta_{\pm} = \frac{3}{2} \pm \sqrt{\frac{9}{4} +m^2}.
\end{align}
As in the standard holographic superconductor model, ${\cal O}_{\phi}^{-}$ and ${\cal O}_{\phi}^{+}$ can be interpreted as the source and the vacuum expectation value of the dual operator, respectively.
In this work we take ${\cal O}_{\phi}^{-}$, the source, to vanish, so that the condensate arises spontaneously.
We note that the $\lambda$ terms in the equation of motion fall off faster than the leading terms near the boundary, and therefore do not contribute to the near-boundary behavior of the scalar field.
 
The third type of solution is characterized by a Lifshitz scaling geometry in the asymptotic region.
Which type a horizon leads to is limited by the Maxwell equation (\ref{eom2}), a first-order equation for the displacement whose source is the charge carried by the scalar.
It fixes $E'$ with the coefficient $\partial_E D$, which vanishes at $E = E_{\rm sat}$ of (\ref{eq:E}), where the displacement stops growing with the field, and is negative above it.
Only horizon data with $E_h < E_{\rm sat}(\phi_h)$ can reach an AdS boundary (section \ref{sec:wall}).
Below a coupling $\lambda^*$, above which no such solution exists (section \ref{sec:lstar}), data beyond this bound can instead flow to the scaling solution
\begin{align}\label{eq:lifansatz}
\phi (r) = \phi_0, ~~~~ A_t (r) = A_0 \, r^z, ~~~~ U(r) = u \, r^2, ~~~~ W(r) = (z-1) \log r,
\end{align}
where $\phi_0$ denotes the value of the scalar field at infinity, and the additive constant of $W$ has been absorbed into $A_0$.
The geometry is then Lifshitz,
\begin{align}
ds^2\big|_{r\rightarrow \infty} \sim - u \, r^{2z} dt^2 + \frac{dr^2}{u \, r^2} + r^2 (dx^2 + dy^2)\,,
\end{align}
where $z$ denotes the dynamical critical exponent of the Lifshitz scaling symmetry.
The field strength $E = A_0 z$ is constant on it and lies above $E_{\rm sat}(\phi_0)$, where $\partial_E D < 0$, and the constants $z$, $A_0$, $\phi_0$ and $u$, through which alone the quartic term enters, are fixed by the algebraic conditions of Appendix \ref{sec:lambdamax}.
Since $D$ is then constant, $r^2 D$ grows like $r^2$ and the scaling solution carries no finite charge density ${\cal Q}$.
It is a fixed point of the radial flow rather than a black hole with a place in the $({\cal Q}/\kappa^2, T/\kappa)$ plane, and the Lifshitz region of the phase diagrams in section~\ref{sec:stages} is the part of the dome without source free solutions, beyond whose edge the horizon data flow to it.
 
\section{Phase diagram and critical point}\label{sec:stages}

\begin{figure}[tbp]
\begin{center}
    \subfigure[$\lambda=0$]{\includegraphics[width=0.42\linewidth]{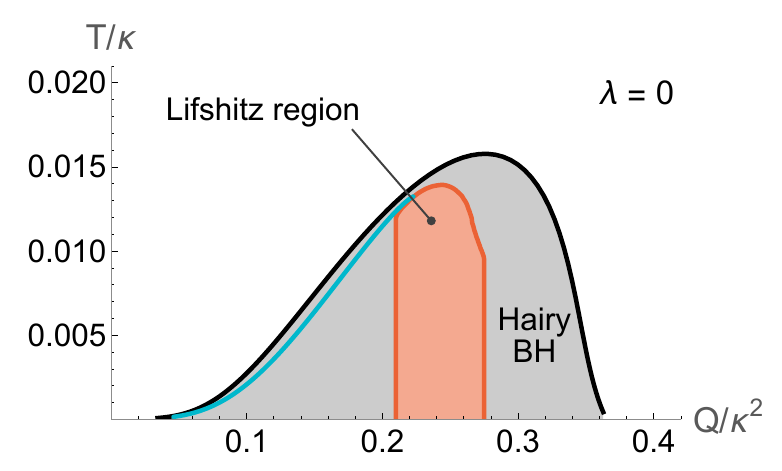}}
    \subfigure[$\lambda=5$]{\includegraphics[width=0.42\linewidth]{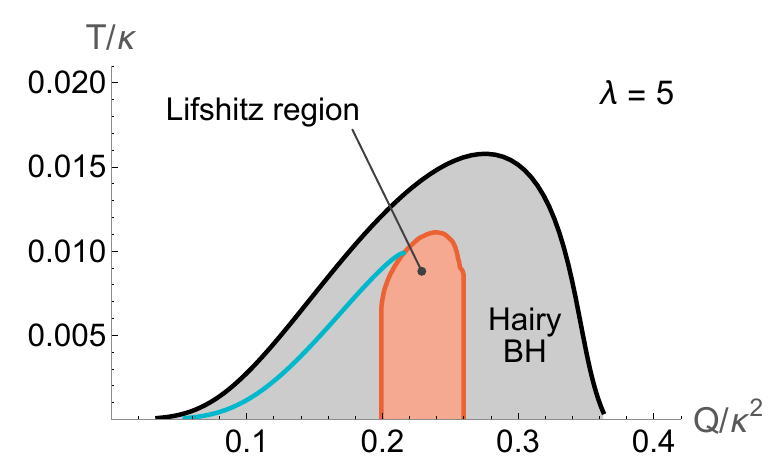}}
    \subfigure[$\lambda=10$]{\includegraphics[width=0.42\linewidth]{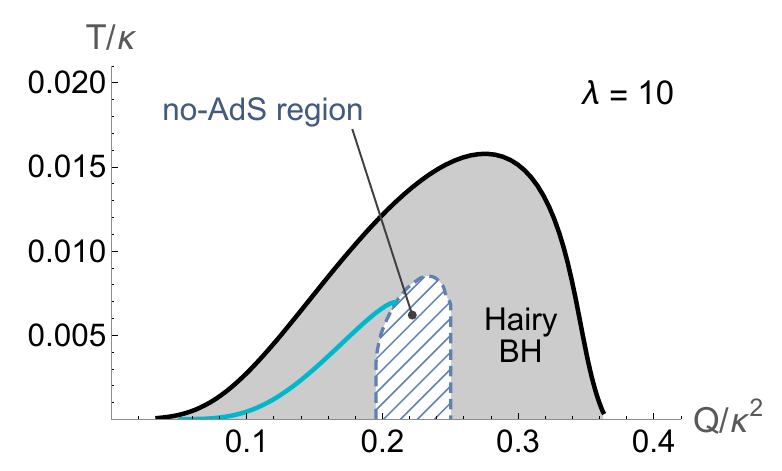}}
    \subfigure[$\lambda=30$]{\includegraphics[width=0.42\linewidth]{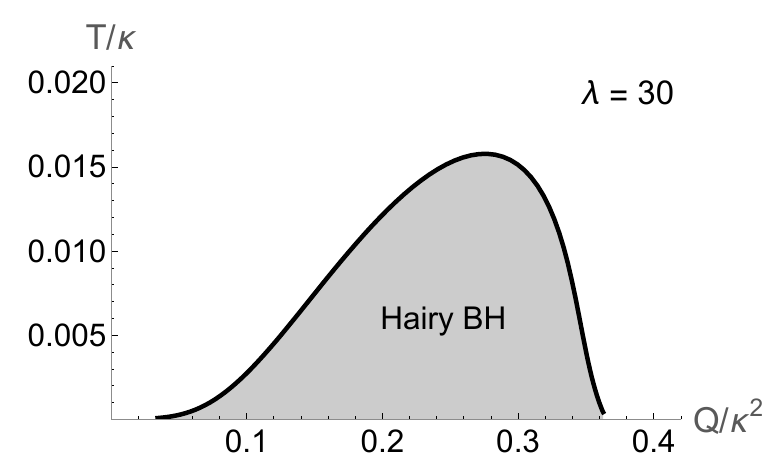}}
    \caption{Phase diagram for (a) $\lambda = 0$, (b) $\lambda = 5$, (c) $\lambda = 10$ and (d) $\lambda = 30$, covering the three stages.
Gray, the hairy black holes, red, the Lifshitz region, hatched blue, the no-AdS region, and cyan, the source free hairy black holes at $\phi_h = 0.22$ of Figure \ref{fig:fatemaps}, none at $\lambda = 30$. Outside the black onset curve the solution is the RN black hole.}
    \label{fig:Phase02}
\end{center}
\end{figure}

The starting point is $\lambda = 0$, the model of Ref.~\cite{Seo:2023kyp}, whose phase diagram for $m^2 = -1$, $q=1$, $\gamma_2=6$ and $\gamma_4=3/2$ is Figure \ref{fig:Phase02}(a).
Outside the dome the solution is the RN black hole.
Owing to the non-linear electrodynamic interaction terms governed by $\gamma_2$ and $\gamma_4$, the hairy black hole solution, corresponding to the superconducting state, occupies a dome-shaped region of the phase diagram (gray shaded region).
At $\lambda = 0$ the transition from the RN black hole to the hairy black holes at the edge of the dome is of second order~\cite{Seo:2023kyp}.
As discussed above, a further region is found inside the dome (red shaded region), bounded above by source free solutions that reach the wall, and beyond its inner edge the flows from the horizon realize the Lifshitz scaling symmetry far from the horizon.
The region is drawn filled down to $T = 0$, although at $\lambda = 0$ the curves of large $\phi_h$ enter its lower part and end there at their folds, and it is the region that the quartic coupling removes in this work.

This section follows the interior of the dome as $\lambda$ is raised, and it contains the main result of this work.
The interior passes through three stages separated by two couplings, $\lambda^*$, at which the Lifshitz fixed point ceases to exist, and $\lambda_c$, above which no source free solution reaches the wall, the field strength $E = E_{\rm sat}$ at which the Maxwell equation degenerates (section \ref{sec:wall}).
Table \ref{tab:regimes} collects them.

\begin{table}[tbp]
\centering
\begin{tabular}{l p{0.36\linewidth} p{0.36\linewidth}}
\hline
 & the deep interior of the dome & beyond its inner edge \\
\hline
$\lambda < \lambda^{*}$
  & a hole, bounded by $E = E_{\rm sat}$ and by a fold of the map to the phase diagram
  & the Lifshitz fixed point of section \ref{sec:lstar} \\
$\lambda^{*} < \lambda < \lambda_{c}$
  & the same hole, still shrinking (the no-AdS region)
  & no asymptotic region, since the flow closes off \\
$\lambda_{c} < \lambda$
  & filled with source free solutions, none of which reaches the wall
  & no inner edge \\
\hline
\end{tabular}
\caption{The three stages of the dome interior, separated by $\lambda^{*}$ of (\ref{eq:lstarvalues}) and $\lambda_c$ of (\ref{eq:lamc}). The hole is the region of the dome that the source free hairy black holes do not fill (section \ref{sec:wall}).}
\label{tab:regimes}
\end{table}

As we increase the parameter $\lambda$ in (\ref{eq:Veffcov}), the phase diagram changes as shown in Figure \ref{fig:Phase02}.
The onset curve, where the hairy black holes branch off the RN black hole, is independent of $\lambda$, since the quartic term is negligible where the scalar is small.
As $\lambda$ grows the upper edge of the Lifshitz region moves down and the region shrinks.

The four panels of Figure \ref{fig:Phase02} are the stages of this section.
In (a) and (b), $\lambda < \lambda^*$, the Lifshitz region is lower at the larger coupling.
In (c), $\lambda^* < \lambda < \lambda_c$, the hole is still there but the flows beyond its inner edge no longer reach a Lifshitz fixed point, and it is hatched as the no-AdS region of section \ref{sec:lstar}.
In (d), $\lambda > \lambda_c$, the hole has closed and the whole dome consists of hairy black holes with an AdS boundary.

\subsection{\texorpdfstring{The wall and the Lifshitz region below $\lambda^*$}{The wall and the Lifshitz region below lambda*}}\label{sec:wall}

Where $E$ reaches $E_{\rm sat}$ of (\ref{eq:E}) the principal part of the Maxwell equation degenerates, so a hairy black hole carrying an asymptotically AdS region exists only while $E < E_{\rm sat}$ throughout the exterior.
On the horizon, where $E = E_h$, the bound is algebraic in the data of (\ref{eq:horizon}),
\begin{align}\label{eq:xh}
E_h < E_{\rm sat} (\phi_h) ,
\end{align}
and neither $\kappa$ nor $\lambda$ enters it.
Horizon data beyond the wall cannot reach an AdS boundary, and the rest of this section is organized by what lies beyond it.

The condensed solutions behind the phase diagrams respect the bound, and those at the edge of the hole reach it.
At the smaller couplings they meet $E = E_{\rm sat}$ at the horizon, where $E_h / E_{\rm sat} (\phi_h)$ comes arbitrarily close to one, and close to $\lambda_c$ they meet it just outside the horizon, with $E_h / E_{\rm sat} (\phi_h)$ below one. No source free solution exceeds the bound (\ref{eq:xh}).\footnote{Near $T = 0$ the horizon values $\phi'(r_h)$ and $W'(r_h)$ of (\ref{eq:horizon}) grow like $1/T$ and $1/T^2$, and $W$ can grow by order one within the first step of the numerical integration away from the horizon.
Horizon data beyond the wall can then start the integration already below it, having stepped over $E = E_{\rm sat}$.
Such horizon data do not give solutions and are discarded, and every one of them starts below the wall at the first integration point.}
Part of the boundary of the source free family is therefore the set of solutions on which $E$ reaches $E_{\rm sat}$, at the horizon or just outside it, and the image of that set in the $({\cal Q}/\kappa^2, T/\kappa)$ plane bounds a region inside the dome that the source free solutions do not fill.
We call it the hole.
The other side of the same region is a fold of the map $(\phi_h, E_h) \mapsto ({\cal Q}/\kappa^2, T/\kappa)$, across which the family continues while its image turns back, so nothing lies beyond that side either.

\begin{figure}[tbp]
    \centering
    \includegraphics[width=0.78\linewidth]{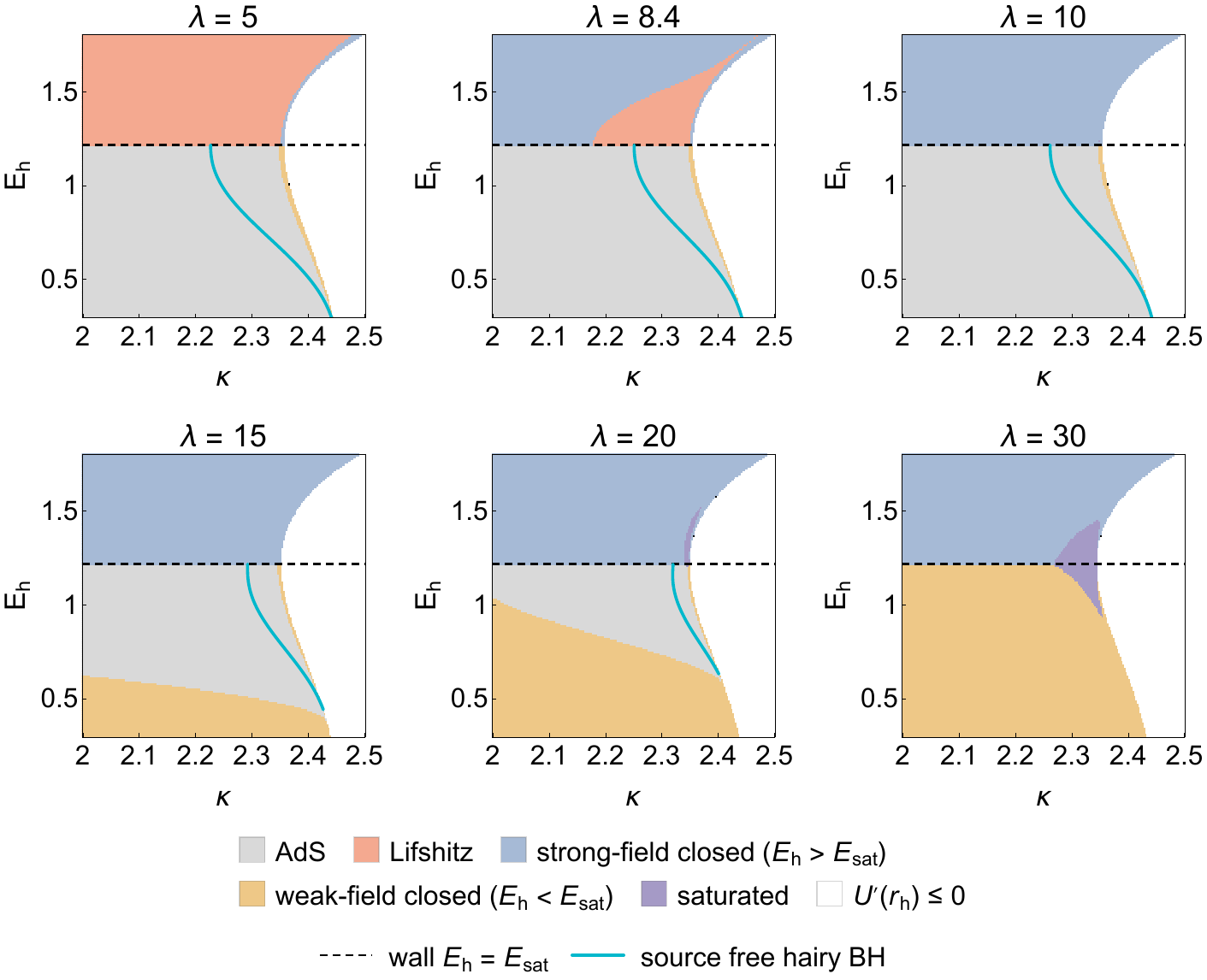}
    \caption{Where the flow from horizon data at $\phi_h = 0.22$ ends, over a window of the plane of $\kappa$ and $E_h$, for six values of $\lambda$.
The dashed line is the wall $E_h = E_{\rm sat} (\phi_h)$, the cyan line the source free hairy black holes, and white the data with $U'(r_h) \le 0$.}
    \label{fig:fatemaps}
\end{figure}

What lies beyond the wall is the fate of the flows from horizon data with $E_h > E_{\rm sat} (\phi_h)$, with no source free condition imposed.
Figure \ref{fig:fatemaps} shows the outcome at $\phi_h = 0.22$ over a window of the plane of $\kappa$ and $E_h$.
For $\lambda < \lambda^*$ a band of horizon data just beyond the wall flows to the Lifshitz fixed point, and this band is the basin of attraction of that fixed point.
Above the band the flow closes off at a finite distance, as described in section \ref{sec:lstar}.
The band narrows as $\lambda$ grows and is absent at $\lambda = 10$.

The color of each cell of Figure \ref{fig:fatemaps} is the fate of the flow from that horizon datum, and the white cells, with $U'(r_h) \le 0$, have no horizon at positive temperature.
Gray cells reach AdS, in general with a source, and the source free hairy black holes at this $\phi_h$ form the cyan line, which ends on the wall.
In the phase diagram of Figure \ref{fig:Phase02} the same solutions are the cyan line, whose end on the wall lies on the edge of the hole.
At fixed $\phi_h$ the wall is a single value of $E_h$, the same for every $\kappa$ and $\lambda$.

Below the wall a second fate without an AdS boundary appears, as a thin strip next to $T = 0$ at smaller $\lambda$ and over a wide range of data at larger $\lambda$, in which $E$ stays well below $E_{\rm sat}$ while the scalar grows away from $\phi_h$ until $U \rightarrow 0$ and $W \rightarrow \infty$ at a finite radius.
This is where the flows beyond the wall stop as well.
Followed past that radius (Appendix \ref{sec:beyondwall}, the orange flow of Figure \ref{fig:blowup}), the flow closes in the same way.
The transverse size turns over, and the flow ends at the fold of $D(E)$, which it reaches only after the turn.
We call a flow that ends this way closed, strong-field closed when it starts beyond the wall, with $E/E_{\rm sat} > 1$ at the horizon, and weak-field closed when it starts below it, the blue and orange cells of Figure \ref{fig:fatemaps}.

Strong-field and weak-field closed flows differ in what keeps the flow from the AdS boundary, not in how the spacetime closes.
Beyond the wall it is the electric field, through the bound (\ref{eq:xh}), whatever $\kappa$ and $\lambda$ are.
Below the wall the electric field allows an AdS boundary, and it is the scalar that fails to decay, which depends on $\lambda$, $\kappa$ and $E_h$.

The weak-field closed flows spread from small $E_h$ and small $\kappa$ as $\lambda$ grows, and at $\lambda = 30$ no horizon data at this $\phi_h$ reach AdS and there is no cyan line, consistent with the source free family ending below it (section \ref{sec:cuts}).
Weak-field closed flows therefore take horizon data out of the source free family, but they open no region of their own in the phase diagram, which is filled at $\lambda = 30$ (Figure \ref{fig:Phase02}).

Close to $T = 0$ a flow can also end without closing, the purple cells of Figure \ref{fig:fatemaps}, which we call saturated.
The scalar barely moves from $\phi_h$ and the transverse size is still growing, but $E$ reaches $E_{\rm sat}$ just outside the horizon and the flow ends there, the purple flow of Figure \ref{fig:blowup}.
The charge of the scalar requires a displacement $D$ that grows outward, while the largest displacement the non-linear electrodynamics can carry, $D$ at $E = E_{\rm sat}$, falls as $\phi$ grows for $\phi^2 < 1/(2\gamma_2)$, as on these flows, and where the two meet, at $E = E_{\rm sat}$ on either branch of $D(E)$, $dE/dr$ diverges and no real field continues the solution.
A saturated flow therefore does not close off like a closed flow, but ends while the space is still opening, again without an asymptotic region.
At the same $E_h$ a slightly higher temperature gives a closed flow instead.

\subsection{\texorpdfstring{The Lifshitz fixed point disappears at $\lambda^*$, leaving the no-AdS region}{The Lifshitz fixed point disappears at lambda*, leaving the no-AdS region}}\label{sec:lstar}

A Lifshitz fixed point is a solution in which the scalar is frozen at a constant $\phi_0$ while $A_t = A_0 r^z$, $U = u\, r^2$ and $e^{W} = r^{z-1}$ scale with powers of $r$, the scaling solution (\ref{eq:lifansatz}) of section \ref{sec:solutions}.
Substituting it into (\ref{eom1}) to (\ref{eom4}) turns the four differential equations into four algebraic conditions on $(\phi_0, A_0, u, z)$, in which $\lambda$ enters as a coefficient and $\kappa$ drops out (Appendix \ref{sec:lambdamax}).

For $\lambda < \lambda^*$ the four conditions have two solutions.
The lower branch attracts the flow from the horizon, while the upper branch, with a much larger $\phi_0$ falling like $\lambda^{-1/2}$, repels it.
As $\lambda$ grows the two branches approach each other, and at $\lambda^*$ they meet and disappear together.
The coupling $\lambda^*$ is therefore the largest value of $\lambda$ along the curve of Lifshitz fixed points, which gives $\lambda^* \simeq 8.61$ (\ref{eq:lstarvalues}).
Since $\kappa$ does not enter, above $\lambda^*$ no Lifshitz fixed point exists for any horizon data.

For $\lambda^* < \lambda < \lambda_c$ the hole is still in the dome, but the Lifshitz fixed point that used to lie beyond its inner edge no longer exists, and the flows from beyond that edge end without any asymptotic region.
We call the hole in this range of couplings the no-AdS region.

Since no Lifshitz fixed point exists above $\lambda^*$, the flows from horizon data beyond the wall $E_h = E_{\rm sat} (\phi_h)$ of (\ref{eq:xh}) instead reach a finite radius $r = r_s$ outside the horizon, where $U \rightarrow 0$ and $W \rightarrow \infty$ and the coordinate $r$ stops growing.
Appendix \ref{sec:beyondwall} shows that $r_s$ is neither a horizon nor a curvature singularity and follows the solution past it, to an end at a finite proper distance from the horizon.
These are therefore genuine local solutions of the bulk equations, but they are not admissible holographic solutions.
The geometry closes off before any asymptotic region is reached, so neither the source ${\cal O}_{\phi}^{-}$ and the condensate ${\cal O}_{\phi}^{+}$ nor $\mu$, ${\cal Q}$ and $T$ can be read off, and no point of the phase diagram corresponds to them.

\begin{figure}[tbp]
    \centering
    \includegraphics[width=0.9\linewidth]{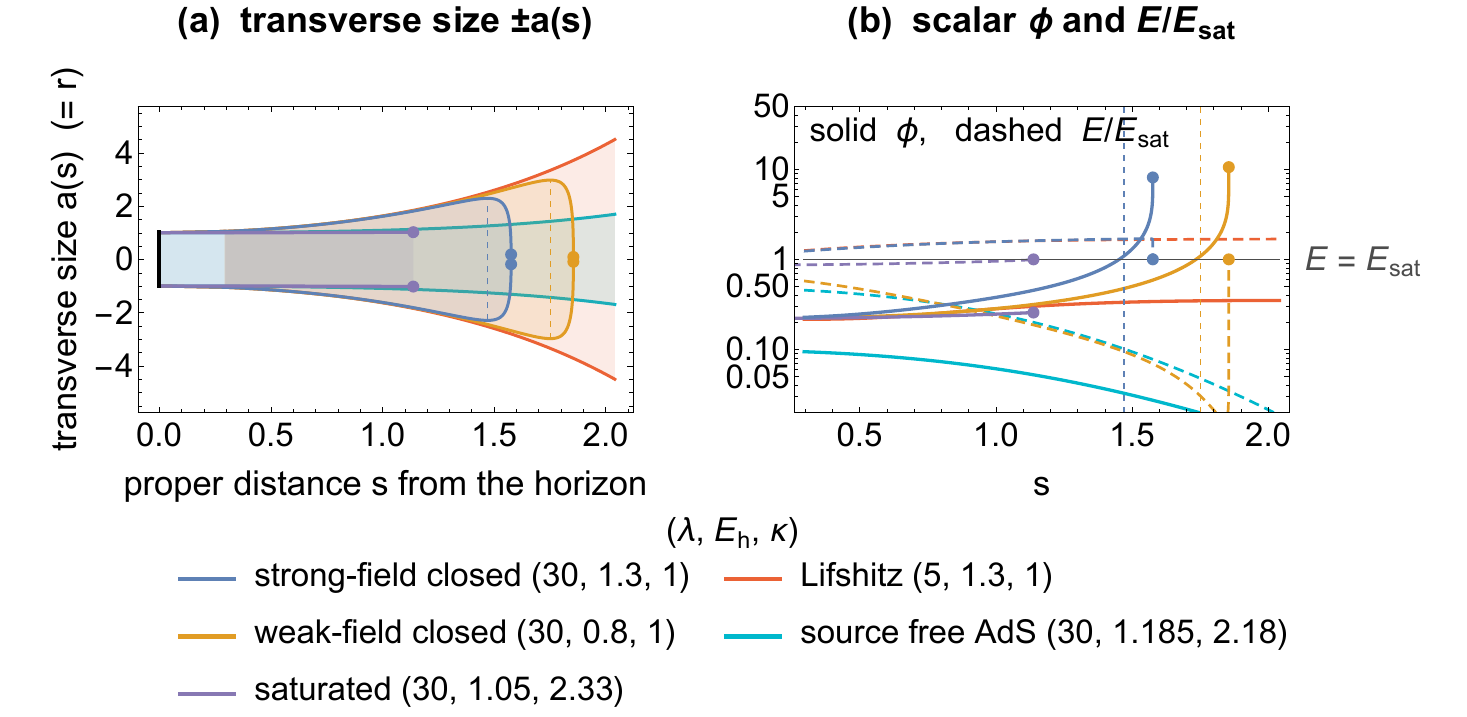}
    \caption{Five flows from the horizon, the non-AdS ones colored as in Figure \ref{fig:fatemaps}, against the proper radial distance $s$  from it, $dr/ds = \sqrt{U}$ (Appendix \ref{sec:beyondwall}).
(a) Transverse size and (b) the scalar (solid) and $E/E_{\rm sat}$ (dashed), with vertical dashed lines at the turning points $s_s$ of (\ref{eq:amax}).}
    \label{fig:blowup}
\end{figure}

The physical meaning of the missing AdS region follows from reading the radial direction as the energy scale of the boundary theory.
The horizon describes the infrared, and the AdS boundary the ultraviolet conformal field theory.
A solution that reaches AdS is an infrared state reached by a renormalization group flow from that conformal field theory, while a solution that ends before reaching AdS is an infrared state with no such ultraviolet completion.
Both AdS and the Lifshitz geometry are fixed points in this sense, since each is invariant under a rescaling of $r$.
AdS, with $z = 1$ and $\phi \rightarrow 0$, is the relativistic conformal field theory, and the Lifshitz geometry, with $z > 1$ and $\phi \rightarrow \phi_0$, is a non-relativistic one.

Which fixed point a solution reaches is decided by its horizon data.
A source free hairy black hole reaches AdS, with $\phi$ decaying as in (\ref{aymphi}).
Horizon data beyond the wall generically cannot reach AdS, because $E / E_{\rm sat}$ would have to pass through one on the way to zero at the boundary, where the Maxwell equation degenerates.
For $\lambda < \lambda^*$ those in the basin of attraction of the Lifshitz fixed point are drawn to it instead, so the flow ends on a different scaling theory, while the rest end as described above.
Figure \ref{fig:blowup} shows the outcomes side by side, and Figure \ref{fig:fatemaps} shows which horizon data end where.

Inside the hole no source free hairy black hole exists, so the only solution with an AdS boundary there is the RN black hole, which is unstable to the condensation that defines the dome.
Within the homogeneous ansatz (\ref{Ansatz}) the model therefore has no stable solution there, and describing it requires either a less symmetric ansatz or a completion of the non-linear electrodynamics, whose response stops growing at $E = E_{\rm sat}$.

\begin{figure}[tbp]
    \centering
    \includegraphics[width=0.62\linewidth]{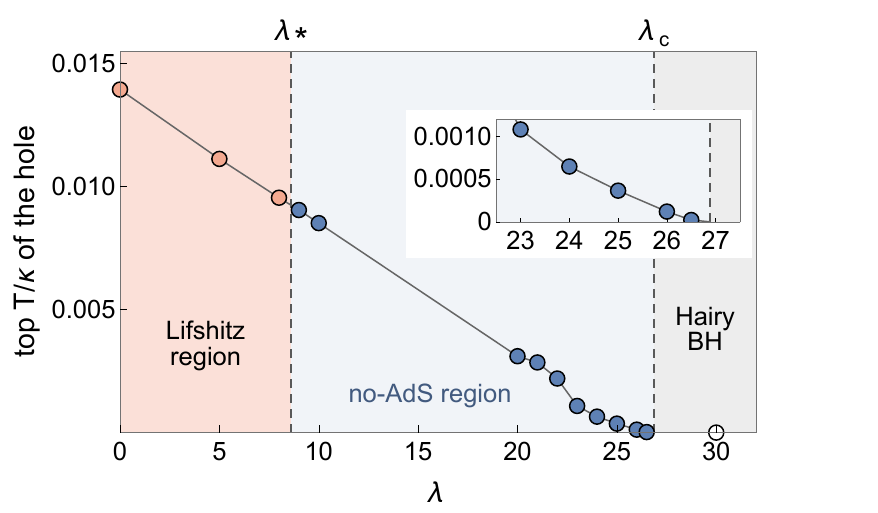}
    \caption{Height of the hole, the largest $T/\kappa$ of the region of the dome that the source free hairy black holes do not fill, against $\lambda$, with $\lambda^*$ and $\lambda_c$ dashed and the three stages shaded.
Light red points mark a hole beyond whose edge the flows reach the Lifshitz fixed point, blue points the no-AdS region and the open point no hole. The inset enlarges the approach to $\lambda_c$, where the line ends.}
    \label{fig:holelambda}
\end{figure}

The no-AdS region still shrinks as $\lambda$ increases (Figure \ref{fig:holelambda}).
It closes at the coupling $\lambda_c$ of section \ref{sec:above}, above which the whole region is covered by hairy black holes with an asymptotically AdS geometry.
In this work we focus on the hairy black holes with an asymptotically AdS boundary, and the Lifshitz fixed point and the flows beyond the wall are discussed in Appendices \ref{sec:lambdamax} and \ref{sec:beyondwall}.

\subsection{\texorpdfstring{Above $\lambda_c$, a filled dome}{Above lambda c, a filled dome}}\label{sec:above}

Raising $\lambda$ shrinks the hole further, and it closes together with the last source free solutions that meet the wall at
\begin{align}\label{eq:lamc}
\lambda_c \simeq 26.9 ,
\end{align}
above which no source free solution meets the wall and the whole dome is filled with source free hairy black holes with an AdS boundary.

\begin{figure}[tbp]
    \centering
    \includegraphics[width=0.70\linewidth]{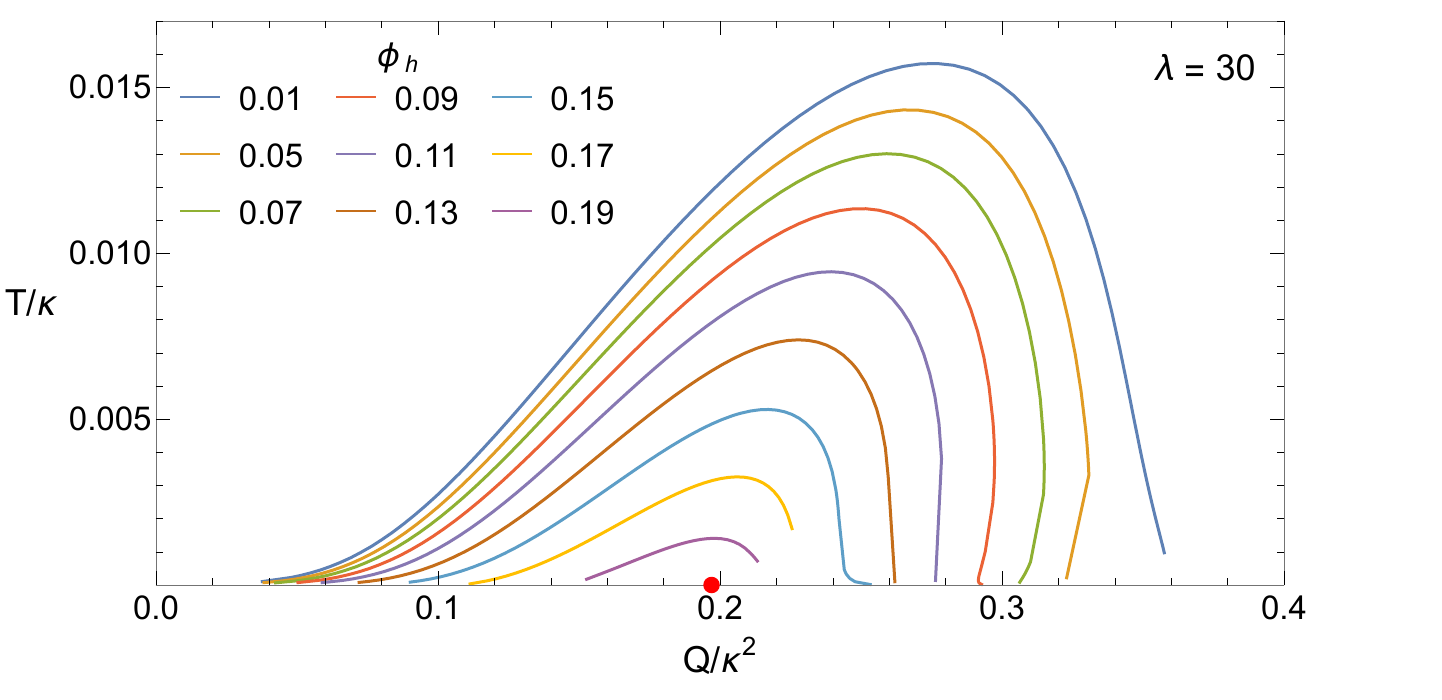}
    \caption{Curves of fixed $\phi_h$ in the dome at $\lambda = 30$, from small $\phi_h$ near the onset curve to large $\phi_h$ deep inside, the curves of large $\phi_h$ ending at folds of the map.
The red dot is the critical point.}
    \label{fig:L30PT}
\end{figure}

The phase diagram at $\lambda = 30$ and the horizon value $\phi_h$ of the scalar field are drawn in Figure \ref{fig:L30PT}.
Outside the curve of smallest $\phi_h$ lies the RN black hole with $\phi_h = 0$.
Inside the superconducting phase, the horizon value of the scalar field $\phi_h$ increases as the temperature decreases.
At the bottom of the dome the curves of fixed $\phi_h$ close in on a point, the red dot in the figure, which we call the ``{\it critical point}".

It lies at ${\cal Q}/\kappa^2 \simeq 0.197$, where the apexes of these curves reach $T = 0$ as $\phi_h$ approaches its largest value (section \ref{sec:cuts}).
From the boundary theory point of view, the condensate grows with $\phi_h$ at fixed charge density, and hence it is large near the critical point.
Each curve in Figure \ref{fig:L30PT} is the upper edge of the source free solutions with one value of $\phi_h$, and at the critical point $\phi_h$ sits at the deepest minimum that the horizon potential can have (section \ref{sec:cuts}).

One interesting observation is that this critical point hardly moves in the phase diagram as $\lambda$ changes above $\lambda_c$.
As $\lambda$ increases the horizon values decrease, but the overall shape of each curve does not change.

The deep interior of the superconducting phase therefore passes through the three stages of Table \ref{tab:regimes}.
Raising $\lambda$ past $\lambda^{*}$ does not change how the hole shrinks, and its height keeps falling at the same rate (Figure \ref{fig:holelambda}).
What changes is what lies beyond it, which is emptied.
The wall stands where it stood, and the Lifshitz fixed point that used to lie beyond it no longer exists, so the Lifshitz region shaded in Figure \ref{fig:Phase02}(a) survives as a region into which no solution of the model continues at all.

Only at $\lambda_c$ does the wall itself move out of reach.
At $\lambda = 30$, the case drawn in Figure \ref{fig:L30PT}, the boundary of the source free family consists of folds of the map and of the extremal $T \rightarrow 0$ edge and of nothing else, and the largest value of $E_h / E_{\rm sat} (\phi_h)$ anywhere on the family stays below one.

Figure \ref{fig:holelambda} follows the height of the hole through all three stages.
It falls monotonically from $\lambda = 0$ to zero at $\lambda_c$, and nothing happens to it at $\lambda^*$, where only what lies beyond its inner edge changes.

\section{The effective potential and the critical point}\label{sec:veff}

\subsection{Boundary and horizon potentials}
In this section we analyze the effective potential of the scalar field in the superconducting phase.
It is the potential that the scalar equation of motion (\ref{eom1}) sees (section \ref{sec:vloc}), namely the potential (\ref{eq:Veffcov}) on the ansatz, written with the field strength (\ref{eq:E}), together with the mass term that the gauge covariant derivative in $|D\varphi|^2$ gives the charged scalar,
\begin{align}\label{eq:Veff}
V_{eff} (r, \phi) = V(\phi, F^2) - \frac{q^2 e^{-2 W} A_t^2 \phi^2}{U} = m_{eff}^2 (r) \, \phi^2 + \lambda \phi^4 ,
\end{align}
with
\begin{align}\label{eq:meff}
m_{eff}^2 (r) = m^2 - \frac{\gamma_2}{2} E^2 + \gamma_4 E^4 - \frac{q^2 e^{-2 W} A_t^2}{U} .
\end{align}

We evaluate (\ref{eq:Veff}) on a solution, with its $A_t$ and $W$, and ask where the scalar of the same solution sits on it.

At the boundary, the field strength $E$ is suppressed and the charge term vanishes, and the effective potential becomes
\begin{align}
V_{eff} (r \rightarrow \infty, \phi) = m^2 \phi^2 + \lambda \phi^4 ,
\end{align}
which for $m^2 < 0$ and $\lambda > 0$ is a double well.
The solutions of this work are source free, so the scalar vanishes at the boundary and sits at the center of this double well, for every solution in the superconducting phase.

At the horizon the effective potential is not universal.
There $E_h$ and $\phi_h$ do not vanish, while the charge term does, because $A_t^2 / U \rightarrow 0$, and in terms of the horizon data (\ref{eq:horizon}) the potential (\ref{eq:Veff}) becomes
\begin{align}\label{eq:Veff_horizon}
V_{eff} (r_h, \phi) = m^2 \phi^2 +\lambda \, \phi^4 -\frac{\phi^2}{2} E_h^2 \left(\gamma_2  -2  \gamma_4 E_h^2 \right) = m_{eff}^2 (r_h) \, \phi^2 + \lambda \phi^4 ,
\end{align}
which is a function of $\phi$ fixed by the horizon datum $E_h$, and the scalar of the solution sits on it at $\phi = \phi_h$.

\begin{figure}[tbp]
    \centering
    \subfigure[]{\includegraphics[height=0.3\linewidth]{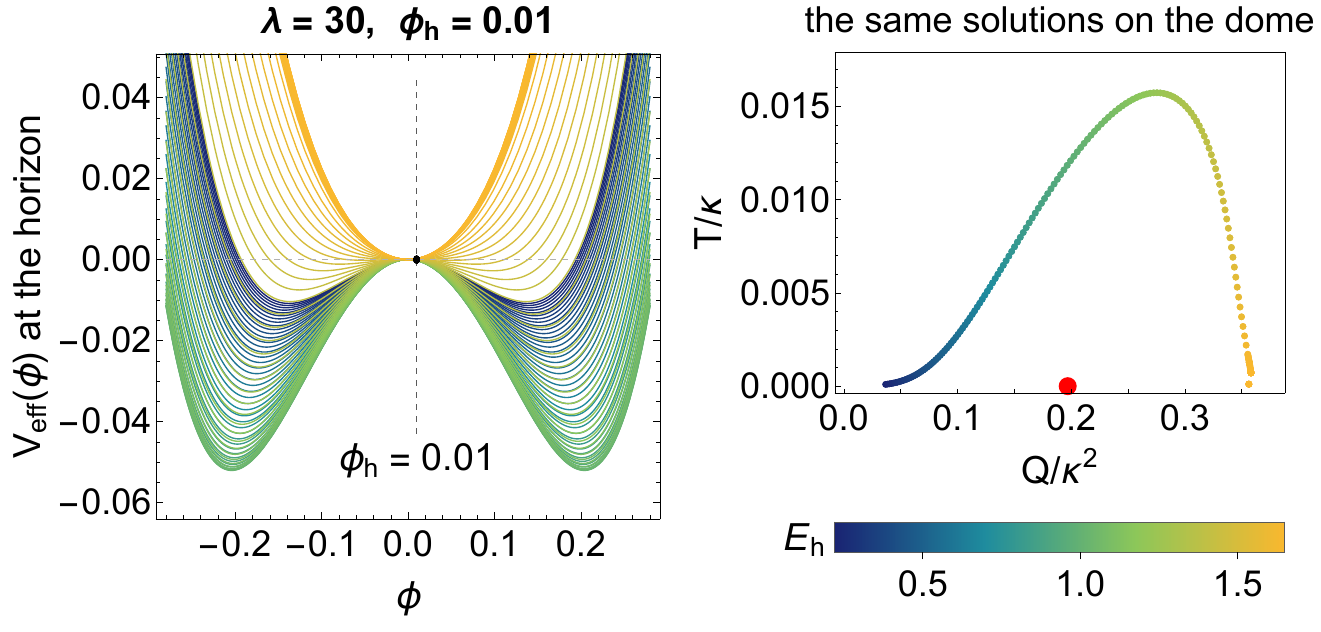}}
    \subfigure[]{\includegraphics[height=0.3\linewidth]{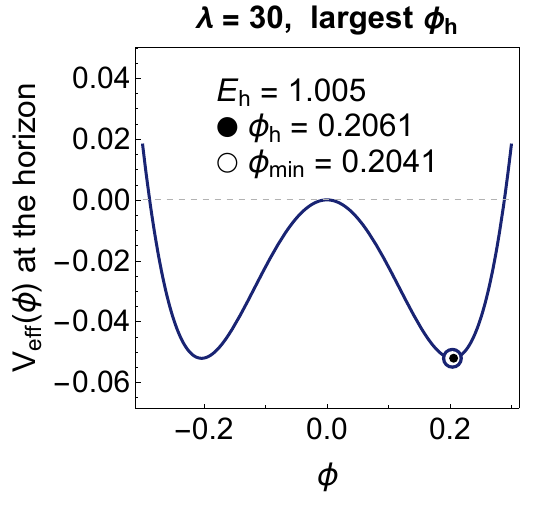}}
    \caption{(a) The horizon effective potential (\ref{eq:Veff_horizon}) of the source free solutions on the upper edge of a dome of small $\phi_h$ at $\lambda = 30$, colored by $E_h$ (left), and the same solutions on the dome (right), with the critical point red.
(b) The horizon potential of the source free solution with the largest $\phi_h$, next to the critical point, with $\phi_h$ (filled) and $\phi_{min}$ (open).}
    \label{fig:Veff_L30_01}
\end{figure}

Figure \ref{fig:Veff_L30_01} shows the horizon effective potential along a curve of small $\phi_h$ near the phase boundary in (a) and at the critical point in (b).
Each color corresponds to the position on that curve drawn in the right panel of Figure \ref{fig:Veff_L30_01}(a).
Along it the horizon potential is a single or a double well, depending on the charge density and the temperature.
Since $\phi_h$, the black dot, is small there, the scalar sits near the origin, at the bottom of a single well or at the top of a double well.

Figure \ref{fig:Veff_L30_01}(b) shows the horizon potential of the source free solution with the largest $\phi_h$, next to the critical point.
Its horizon value lies at the minimum of the well, and this well is the deepest one that the horizon potential can have (section \ref{sec:cuts}).
The same holds for every $\lambda$ above $\lambda_c$, and the next two subsections explain why.

\subsection{The horizon identity}\label{sec:vloc}

That (\ref{eq:Veff}) is the potential the scalar sees follows from multiplying (\ref{eom1}) by $U$ and collecting the terms that carry no derivative of $\phi$,
\begin{align}\label{eq:flux}
U \phi '' + \left[ U' + U \left( \frac{2}{r} + W' \right) \right] \phi ' = \frac{1}{2} \frac{\partial V_{eff}}{\partial \phi}.
\end{align}
The last term of (\ref{eq:meff}) comes from $|D\varphi|^2$, in which the gauge covariant derivative contributes $q^2 g^{tt} A_t^2 |\varphi|^2$, which is a mass term for $\phi$ with an $r$-dependent, negative coefficient, since $g^{tt} < 0$.
It is this term that drives the condensation, and it is what distinguishes $V_{eff}$ from $V$.

At a black hole horizon both $U$ and $A_t$ vanish, and (\ref{eq:flux}) degenerates, because the coefficient of the highest derivative goes to zero.
Requiring $\phi''(r_h)$ to stay finite kills the first term on the left, while $A_t^2 / U \sim (r - r_h)$ removes the gauge piece on the right, so the charge term drops out of $m_{eff}^2$ and what survives is
\begin{align}\label{eq:hor-identity}
2 \, U'(r_h) \, \phi '(r_h) = \phi_h \left( 2 m^2 - \gamma_2 \, E_h^2 + 2 \gamma_4 \, E_h^4 + 4 \lambda \phi_h^2 \right) = V_{eff}'(\phi_h),
\end{align}
where the bracket is $2 ( m_{eff}^2 (r_h) + 2 \lambda \phi_h^2 )$ with $m_{eff}^2 (r_h)$ of (\ref{eq:Veff_horizon}).
Equation (\ref{eq:hor-identity}) is not new information, since it is the horizon condition (\ref{eq:horizon}) for $\phi'(r_h)$ multiplied by $2 U'(r_h)$.
The factor $U'(r_h)$, which carries the temperature, therefore cancels in the product, and the product depends on $(\phi_h, E_h)$ alone.
In particular the axion charge $\kappa$, which enters $U'(r_h)$ and $\phi'(r_h)$ separately, drops out.
Since $U'(r_h) > 0$ at nonzero temperature, the horizon value $\phi_h$ sits at a stationary point of $V_{eff}$ exactly when the scalar profile starts flat at the horizon, $\phi'(r_h) = 0$, and otherwise the sign of $\phi'(r_h)$ is the sign of $V_{eff}'(\phi_h)$.
As $T \rightarrow 0$ the factor $U'(r_h)$ vanishes, so on solutions whose $\phi'(r_h)$ stays bounded the identity drives $\phi_h$ to a stationary point of the horizon potential.

Whether that extremum is a minimum is decided by the sign of $m_{eff}^2 (r_h)$.
For $m_{eff}^2 < 0$ the origin is a local maximum and $V_{eff}$ has a symmetry-breaking minimum at
\begin{align}\label{eq:phimin}
\phi_{min} = \sqrt{ - \frac{m_{eff}^2 (r_h)}{2 \lambda} } ,
\end{align}
while for $m_{eff}^2 > 0$ the only minimum is $\phi = 0$.
The sign changes at
\begin{align}\label{eq:ahcrit}
E_h^2 = \frac{\gamma_2 + \sqrt{\gamma_2^2 - 16 \gamma_4 m^2}}{4 \gamma_4} ,
\end{align}
which for $m^2 = -1$, $\gamma_2 = 6$ and $\gamma_4 = 3/2$ is $E_h \simeq 1.51$.
Above that field strength the double well has closed up.
This is a different condition from the wall (\ref{eq:xh}), since (\ref{eq:ahcrit}) is a single number while the $E_h$ that saturates (\ref{eq:xh}) moves with $\phi_h$.
The colors of Figure \ref{fig:Veff_L30_01}(a) run across this value.
On a curve of small $\phi_h$ the horizon value sits near the origin, which is the top of a double well for small $E_h$ and the bottom of a single well above (\ref{eq:ahcrit}).
These solutions are condensed although their horizon potential is a single well, so the sign of $m_{eff}^2(r_h)$ does not decide whether the scalar condenses. It is the charge term, which vanishes at the horizon but not outside it, that drives the condensation.

\subsection{Cuts through the dome}\label{sec:cuts}

Stationarity at the horizon is an algebraic condition on the horizon data alone.
Setting $V_{eff}'(\phi_h) = 0$ in (\ref{eq:hor-identity}) gives $\phi_h = \phi_{min}(E_h)$ through (\ref{eq:phimin}) and (\ref{eq:meff}), one relation between $\phi_h$ and $E_h$.
Imposing in addition that the solution be source free, ${\cal O}_{\phi}^{-} = 0$, removes one more freedom from $(\phi_h, E_h, \kappa)$ and leaves a one-parameter family.
The locus on which the horizon value sits exactly at the minimum of $V_{eff}$ is therefore a curve in the $({\cal Q}/\kappa^2, T/\kappa)$ plane, not a region.

At $\lambda = 30$ this curve starts on the $T = 0$ edge to the left of the critical point and runs along it up to about the critical point.
There it turns upward and crosses the interior diagonally towards the right flank of the dome.
Along it $E_h$ grows towards (\ref{eq:ahcrit}) and $\phi_{min}$ falls to zero, so the curve ends on the phase boundary, where $\phi_h \rightarrow 0$.
It divides the dome in two.
Above and to the left of it $V_{eff}'(\phi_h) < 0$ and $\phi'(r_h) < 0$, so $\phi_h$ stops short of the minimum, which covers the whole upper part of the dome.
Below and to the right $\phi_h$ lies beyond the minimum, in the strip between the curve and the low temperature part of the right flank.

\begin{figure}[tbp]
    \centering
    \includegraphics[width=0.6\linewidth]{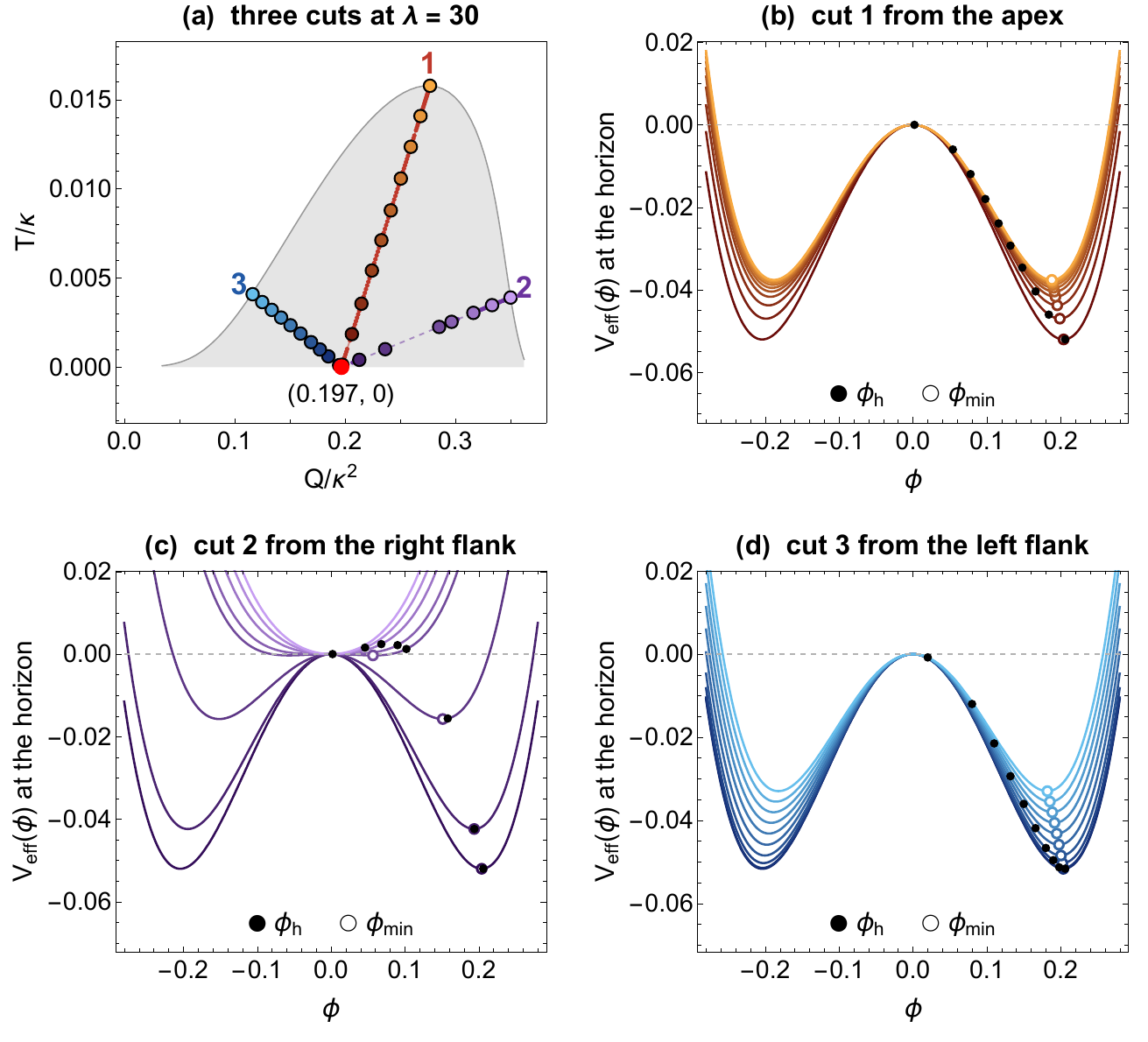}
    \caption{Three cuts through the $\lambda = 30$ dome into the critical point (a), and the horizon effective potential along each, (b) to (d), with $\phi_h$ (filled) and $\phi_{min}$ (open).
In (a) the small dots are the source free solutions on each cut and the large dots those whose potentials are drawn, darker near the critical point (red).}
    \label{fig:cuts}
\end{figure}

The critical point is the point onto which the curves of fixed $\phi_h$ in Figure \ref{fig:L30PT} shrink as $\phi_h$ approaches its largest value.
There $\phi_h = \phi_{min}$ at the deepest minimum of the horizon potential, which the next paragraph identifies, and the locus passes through it.

The same well also sets where the family ends in $\phi_h$.
Over $E_h$ the minimum (\ref{eq:phimin}) is deepest where $m_{eff}^2(r_h)$ is most negative, at $E_h = 1$, so that $\phi_{min} \le \sqrt{5/(4\lambda)}$.
The largest $\phi_h$ that still has a source free solution reaches this bound, and the last solutions lie close to $T = 0$ with $E_h$ close to one.
The top of the family therefore lies at the largest minimum that the horizon potential can have, and it is the critical point.
This is not a bound on individual solutions, many of which have $\phi_h$ beyond the minimum of their own well.

A large fraction of the source free solutions have $V_{eff}'(\phi_h) > 0$, that is $\phi_h$ beyond the minimum and $\phi'(r_h) > 0$.
The sign of $\phi'(r_h)$ therefore does not decide whether a flow below the wall reaches AdS or closes (section \ref{sec:wall}).
At $\phi_h = 0.22$ in Figure \ref{fig:fatemaps} nearly all flows that leave the horizon with $\phi'(r_h) > 0$ fail to reach AdS, but part of the weak-field closed flows leave it with $\phi'(r_h) < 0$ and turn around further out.
At $\lambda = 30$ this $\phi_h$ lies above $\sqrt{5/(4\lambda)}$, which makes $\phi'(r_h) > 0$ for every $E_h$, and no flow below the wall reaches AdS, consistent with the family ending below this $\phi_h$.

How the horizon value reaches the minimum can be seen by cutting the dome along straight lines into the critical point, Figure \ref{fig:cuts}.
Along each of three lines through the critical point $({\cal Q}/\kappa^2, T/\kappa) = (0.197, 0)$, aimed at the apex of the dome, at its right flank and at its left flank, the ratio $\phi_h / \phi_{min}$ runs to unity at the critical point, here at $\lambda = 30$.

Along cuts 1 and 3 the ratio falls towards zero away from the critical point, which is the horizon value sliding back towards the local maximum at the origin as the dome boundary is approached.
Along cut 2 $\phi_h$ stays beyond $\phi_{min}$, in the strip below the locus $\phi'(r_h) = 0$, and near the right flank $E_h$ passes (\ref{eq:ahcrit}) and the double well closes up.
In Figure \ref{fig:cuts} the potentials along each cut are darker nearer to the critical point, and away from it $\phi_h$ moves off $\phi_{min}$ while $\phi'(r_h)$ moves off zero, as (\ref{eq:hor-identity}) requires.

\subsection{The Lifshitz fixed point sits at a minimum as well}\label{sec:lifmin}

The same condition characterizes the Lifshitz fixed point, and there it holds identically.
A frozen scalar has $\phi' = \phi'' = 0$, so (\ref{eq:flux}) reduces to $\partial V_{eff} / \partial \phi = 0$ with no reference to a horizon.
The gauge term does not drop out this time, because $A_t$ does not vanish.
What saves the argument is that on the scaling solution (\ref{eq:lifansatz}) both invariants that enter $m_{eff}^2$ are independent of $r$, $E = A_0 z$ and $e^{-2W} A_t^2 / U = A_0^2 / u$, so
\begin{align}\label{eq:Vlif}
V_{eff} (\phi) = m_{eff}^2 \phi^2 + \lambda \phi^4, ~~~~~~ m_{eff}^2 = m^2 - \frac{\gamma_2}{2} A_0^2 z^2 + \gamma_4 A_0^4 z^4 - \frac{q^2 A_0^2}{u},
\end{align}
is a single fixed double well and $\phi_0$ is its minimum, $\phi_0 = \sqrt{-m_{eff}^2 / 2 \lambda}$.

Below $\lambda^*$ the solution comes in the two branches of section \ref{sec:lstar}.
On both branches $m_{eff}^2 < 0$ for every $0 < \lambda < \lambda^*$, and $\phi_0$ equals $\sqrt{-m_{eff}^2 / 2\lambda}$.
On the lower branch, the one that attracts the flow from the horizon, dropping the charge term is not a small error, since the derivative of $V$ alone at the same solution is of order one instead of zero.
There $\phi_0$ changes little up to $\lambda^*$, so the depth of the well, $\lambda \phi_0^4$, grows nearly in proportion to $\lambda$.

The quartic term turns the potential seen by the scalar into a Landau--Ginzburg double well, $m_{eff}^2 \phi^2 + \lambda \phi^4$, whose quadratic coefficient is set by the electric field and, away from the horizon, by the charge of the scalar, as the quadratic coefficient of the Landau--Ginzburg free energy is set by the temperature.
The Lifshitz fixed point sits exactly at the minimum of its well for every $\lambda < \lambda^*$.
In the filled dome the horizon value sits at the minimum of the horizon well only on the locus $\phi'(r_h) = 0$, and that locus runs along the bottom of the dome and passes through the critical point.
Elsewhere the horizon value is off the minimum, and the slope of the well there is matched by a nonzero $\phi'(r_h)$ through (\ref{eq:hor-identity}).

\section{Discussion}\label{sec:discussion}
In this work, we have investigated a Landau--Ginzburg extension of a holographic superconductor with non-linear electrodynamic interactions by introducing a quartic scalar self-interaction.
The disappearance of the Lifshitz region proceeds in distinct stages.
As $\lambda$ increases, this region shrinks and loses its scaling solution at $\lambda^*$.
Between $\lambda^*$ and $\lambda_c$ a no-AdS region remains, beyond whose edge the flows end before reaching any asymptotic region.
For $\lambda > \lambda_c$, no source free solution reaches the wall, and the dome consists entirely of source free, asymptotically AdS hairy black holes.
At the bottom of the dome the horizon value of the scalar is largest at a point that we refer to as the critical point, which the curves of fixed $\phi_h$ reach as $T \rightarrow 0$.
There the horizon value lies at the deepest symmetry-breaking minimum of the horizon effective potential.

The double-well potential discussed in
section~\ref{sec:veff} is a bulk effective potential evaluated
at the horizon. It is not the Landau--Ginzburg free energy
of the boundary order parameter, and the horizon scalar value
$\phi_h$ should not be identified with the condensate
${\cal O}_{\phi}^{+}$, although at fixed charge density the condensate grows with
$\phi_h$. The association of the critical
point with a minimum of this potential therefore does not,
by itself, establish the dynamical stability of the corresponding bulk solution.
Whether the boundary free energy near this point admits
a Landau--Ginzburg
description remains an open question~\cite{Yin:2013fwa,Bu:2021clf,Natsuume:2024ril,Natsuume:2024sic}. Likewise, the term ``critical point''
is used here to denote the point at the bottom of the dome with the largest
horizon value of the scalar field. It does not, by itself, imply a diverging
correlation length or establish quantum criticality.

On the horizon potential of a condensed solution, the symmetric point $\phi = 0$ is where an uncondensed horizon would sit, and at the critical point the horizon value lies at the deepest symmetry-breaking minimum.
Depending on the parameters, the symmetric extremum can be a local maximum or a local minimum, and near the right flank of the dome the scalar condenses even where it is a minimum (section~\ref{sec:vloc}).
This comparison provides a potential-based characterization of positions on one potential, rather than a description of a real-time evolution, and the normal phase at the same charge density and temperature has different horizon data.
By contrast, for the Lifshitz fixed point discussed in section~\ref{sec:lifmin}, the constant scalar value lies at the minimum of the effective potential (\ref{eq:Vlif}) in its asymptotic scaling region, as the scalar equation requires of a constant scalar.
These observations highlight the distinct roles of the horizon and asymptotic effective potentials in characterizing the superconducting solutions.

Two further limits of the analysis should be kept in mind.
On the Lifshitz fixed point and on the flows from horizon data beyond the wall the field exceeds $E_{\rm sat}$, where $\partial_E D < 0$ and the response of the non-linear electrodynamics to a change of the field is negative, and whether such backgrounds are stable against fluctuations of the gauge field is not settled here.
Finally, the critical point is reached as a limit of solutions at nonzero temperature, and the zero-temperature solutions themselves, whose infrared is expected to be of the kind found in Ref.~\cite{Gubser:2009cg}, are not constructed.

Another notable result is that, for $\lambda > \lambda_c$,
the critical point stays at the bottom of the dome at nearly the same charge
density as $\lambda$ grows. As $\lambda$ increases,
the scalar value at the symmetry-breaking minimum of the
horizon effective potential decreases. The horizon scalar
value at the critical point decreases accordingly and
continues to lie at this minimum. The horizon field at the critical point is set by the effective mass alone, $E_h \simeq 1$ where $m_{eff}^2(r_h)$ is most negative, which does not involve $\lambda$, whereas the quartic coupling sets the scalar amplitude there, $\phi_{min} = \sqrt{5/(4\lambda)}$.

An interesting direction for future work is to apply this
extension to holographic models calibrated against experimental
superconducting domes, following the data-driven
approach of Refs.~\cite{Kim:2024ndba,Kim:2026mll}.
Introducing a quartic scalar self-interaction into such
models would allow us to investigate whether a zero-temperature
state with a maximal horizon value of the scalar field emerges and how its charge-density
location compares with that of an experimentally identified
quantum critical point. Any such correspondence would require
an independent analysis of quantum critical behavior, beyond
the maximization of the horizon value of the scalar field. Nevertheless, this
comparison could help clarify whether the horizon
effective-potential picture captures features relevant
to quantum criticality in unconventional superconductors.

\acknowledgments

We thank Sang-Jin Sin, Kyung Kiu Kim and Johanna Erdmenger for useful discussion.
S.K. is supported by a KIAS Individual Grant (AP103401) via the Center for Artificial Intelligence and Natural Sciences at Korea Institute for Advanced Study.
Y.S. is supported by the National Research Foundation of Korea (NRF) grant funded by the Korea government (MSIT) (No. RS-2026-25588089).
The code for the computations reported here, and drafts of this manuscript, were developed with the assistance of Claude (Anthropic).
The formulation of the problem, the design of the computations and the interpretation of the results are the authors' own, and the authors take full responsibility for the content.
We thank APCTP, Pohang, Korea for their hospitality during the Focus Program [APCTP-2026-T04] from which this work greatly benefited.
As a member of Korea Research Network for Theoretical Physics, we thank the other members for helpful(constructive) discussion.

\appendix
\section{The largest quartic coupling admitting a Lifshitz fixed point}\label{sec:lambdamax}

Whether a Lifshitz fixed point exists at a given $\lambda$ is settled by algebra, before any differential equation is solved.
The fixed point is scale covariant, since $r \rightarrow b r$ together with $t \rightarrow b^{-z} t$ and $x \rightarrow b^{-1} x$ leaves it invariant, and that is what ties every field to a power of $r$ as in (\ref{eq:lifansatz}).
The scalar freezes because any $r$ dependence would break the scaling, and $e^{W} = r^{z-1}$ is what gives time the weight $z$.
The axion enters only through $\kappa^2 / r^2$, which falls off faster than the other terms, so $\kappa$ drops out of the problem entirely.

Substituting (\ref{eq:lifansatz}) into (\ref{eom1}) to (\ref{eom4}) leaves four residues, each a pure power of $r$ times a constant apart from the axion term, so for $\kappa = 0$ the powers of $r$ cancel identically and this is an exact solution, and for $\kappa \neq 0$ it is the fixed point that the flow approaches as $r \rightarrow \infty$.
With $E_0 \equiv A_0 z$ the constant field strength, the four conditions are
\begin{align}\label{eq:lif1}
&m^2 - \frac{\gamma_2}{2} E_0^2 + \gamma_4 E_0^4 - \frac{q^2 A_0^2}{u} + 2 \lambda \phi_0^2 = 0 , \\
\label{eq:lif2}
&u z \left( 1 + \gamma_2 \phi_0^2 - 4 \gamma_4 \phi_0^2 E_0^2 \right) = q^2 \phi_0^2 , \\
\label{eq:lif3}
&12 u = 12 - 2 m^2 \phi_0^2 - 2 \lambda \phi_0^4 - \left( 1 + \gamma_2 \phi_0^2 \right) E_0^2 + 6 \gamma_4 \phi_0^2 E_0^4 - \frac{2 q^2 A_0^2 \phi_0^2}{u} , \\
\label{eq:lif4}
&2 u^2 (z-1) = q^2 A_0^2 \phi_0^2 ,
\end{align}
from the scalar equation, the Maxwell equation, the equation for $U$ and the equation for $W$.
The first is the statement that $\phi_0$ sits at the minimum of (\ref{eq:Vlif}), and the last fixes $A_0$ once the additive constant of $W$ is set to zero as in (\ref{eq:lifansatz}).

Three of the unknowns follow from $\phi_0$ and $E_0$ in closed form.
Writing $\varepsilon \equiv D / E = 1 + \gamma_2 \phi_0^2 - 4 \gamma_4 \phi_0^2 E_0^2$ for the displacement (\ref{eq:E}) divided by the field, (\ref{eq:lif2}) and (\ref{eq:lif4}) give
\begin{align}\label{eq:lifsolved}
z = 1 + \frac{E_0^2 \varepsilon^2}{2 q^2 \phi_0^2} , ~~~~~~ u = \frac{q^2 \phi_0^2}{z \varepsilon} , ~~~~~~ A_0 = \frac{E_0}{z} ,
\end{align}
and (\ref{eq:lif1}) then gives the coupling itself,
\begin{align}\label{eq:liflambda}
\lambda = - \frac{m_{eff}^2}{2 \phi_0^2} , ~~~~~~ m_{eff}^2 = m^2 - \frac{\gamma_2}{2} E_0^2 + \gamma_4 E_0^4 - \frac{E_0^2 \varepsilon}{z \phi_0^2} .
\end{align}
Substituting (\ref{eq:lifsolved}) and (\ref{eq:liflambda}) into (\ref{eq:lif3}) and clearing denominators leaves a single polynomial condition $P(\phi_0^2, E_0^2) = 0$, of degree four in $\phi_0^2$ and six in $E_0^2$,
\begin{align}\label{eq:Gpoly}
P &= 2 E_0^4 - 24 E_0^2 \nonumber\\
&\quad + \phi_0^2 \Big[ - 48 q^2 + \left( 2 m^2 + 16 q^2 - 72 \gamma_2 \right) E_0^2 + \left( 9 \gamma_2 + 288 \gamma_4 \right) E_0^4 - 38 \gamma_4 E_0^6 \Big] \nonumber\\
&\quad + \phi_0^4 \Big[ 48 q^4 + 4 m^2 q^2 - 48 \gamma_2 q^2 + \left( 6 \gamma_2 m^2 + 34 \gamma_2 q^2 + 192 \gamma_4 q^2 - 72 \gamma_2^2 \right) E_0^2 \nonumber\\
&\qquad\quad + \left( 15 \gamma_2^2 + 576 \gamma_2 \gamma_4 - 24 \gamma_4 m^2 - 140 \gamma_4 q^2 \right) E_0^4 \nonumber\\
&\qquad\quad - \left( 126 \gamma_2 \gamma_4 + 1152 \gamma_4^2 \right) E_0^6 + 264 \gamma_4^2 E_0^8 \Big] \nonumber\\
&\quad + \phi_0^6 \left( \gamma_2 - 4 \gamma_4 E_0^2 \right) \Big[ 4 m^2 q^2 + \left( 6 \gamma_2 m^2 + 18 \gamma_2 q^2 - 24 \gamma_2^2 \right) E_0^2 \nonumber\\
&\qquad\quad + \left( 11 \gamma_2^2 + 192 \gamma_2 \gamma_4 - 24 \gamma_4 m^2 - 76 \gamma_4 q^2 \right) E_0^4 \nonumber\\
&\qquad\quad - \left( 94 \gamma_2 \gamma_4 + 384 \gamma_4^2 \right) E_0^6 + 200 \gamma_4^2 E_0^8 \Big] \nonumber\\
&\quad + \phi_0^8 E_0^2 \left( \gamma_2 - 4 \gamma_4 E_0^2 \right)^3 \left( 2 m^2 + 3 \gamma_2 E_0^2 - 14 \gamma_4 E_0^4 \right) .
\end{align}
The Lifshitz fixed points therefore form a curve in the plane of $(\phi_0^2, E_0^2)$, and (\ref{eq:liflambda}) is an explicit function on it.

The coupling $\lambda^*$ is the largest value of $\lambda$ on this curve.
For $\lambda < \lambda^*$ a line of constant $\lambda$ meets the curve twice at $z < 2$, once on the branch that attracts the flow from the horizon and once on the branch that repels it, and at $\lambda^*$ the two points merge.
At the maximum the gradient of $\lambda$ is parallel to that of $P$,
\begin{align}\label{eq:lif5}
P = 0 , ~~~~~~ \frac{\partial (\lambda, P)}{\partial (\phi_0^2, E_0^2)} = 0 ,
\end{align}
two equations for two unknowns.
For $m^2 = -1$, $\gamma_2 = 6$, $\gamma_4 = 3/2$ and $q = 1$ their solution gives
\begin{align}\label{eq:lstarvalues}
\lambda^* &= 8.614515, \cr
\phi_0 &= 0.382858, \cr
A_0 &= 1.281580, \cr
u &= 1.082261, \cr
z &= 1.102771 .
\end{align}
The coupling $\lambda^*$ is an algebraic number.
For these parameters it is a root of an irreducible polynomial of degree $23$ with integer coefficients, and (\ref{eq:lif5}) is the most explicit statement of it.

\section{Flows beyond the wall}\label{sec:beyondwall}

This appendix describes the flows of section \ref{sec:lstar}, which start from horizon data beyond the wall and reach a finite radius $r_s$ at which the coordinate $r$ stops growing.
The horizon data obey the regularity conditions (\ref{eq:horizon}), so the horizon itself is regular.

Near $r_s$ the solutions behave as
\begin{align}\label{eq:rsnear}
U(r) \simeq U_1 \left( r_s - r \right), ~~~~~~ W(r) \simeq -\frac{1}{2} \ln \left( r_s - r \right),
\end{align}
with a positive constant $U_1$.
The divergence of $W$ cancels the zero of $U$ in $g_{tt} = -U e^{2W}$, which stays finite, and the curvature stays finite as well, so $r_s$ is neither a horizon nor a curvature singularity.
What fails is the coordinate.
In the metric (\ref{Ansatz}) $r$ is the area radius of the transverse plane, and with $s$ the proper radial distance,
\begin{align}\label{eq:drds}
\frac{dr}{ds} = \sqrt{U} \rightarrow 0 \quad (r \rightarrow r_s),
\end{align}
so $r_s$ is where the transverse area stops growing.

Past $r_s$ the solution is continued in the gauge
\begin{align}\label{eq:propgauge}
g_{MN}\, dx^M dx^N = -f(s)\, dt^2 + ds^2 + a(s)^2 \left( dx^2 + dy^2 \right), ~~~~~~ a = r, ~~~ f = U e^{2W},
\end{align}
in which nothing is special at $r_s$.
Integrating (\ref{eq:drds}) with (\ref{eq:rsnear}) gives $r_s - r \propto (s - s_s)^2$, so the transverse size has a smooth maximum,
\begin{align}\label{eq:amax}
a(s) \simeq r_s - \frac{U_1}{4} \left( s - s_s \right)^2 ,
\end{align}
where $s_s$ is the proper distance of the turning point.
Beyond it the transverse space contracts while the scalar grows.

The Maxwell equation (\ref{eom2}) fixes $E'$ through $\partial_E D$, with the displacement (\ref{eq:E}), and cannot be solved for $E'$ once $E \rightarrow E_{\rm sat}$.
This is the fold of $D(E)$, where the response of the non-linear electrodynamics stops growing with the field.
Along the contracting flows the scalar becomes large, and for large $\phi$ the fold sits at
\begin{align}\label{eq:Elimit}
E^2 = E_{\rm sat}^2 (\phi) \rightarrow \frac{\gamma_2}{12 \gamma_4} = \frac{1}{3} .
\end{align}
The flow ends in one of two ways.
Either it reaches the fold, $E \rightarrow E_{\rm sat}(\phi)$, which tends to $1/\sqrt{3}$ once the scalar is large, at a finite proper distance, with $E - E_{\rm sat} \propto \sqrt{s_{end} - s}$ and a curvature that is large but finite, or the transverse space collapses, $a \rightarrow 0$, and the curvature diverges.

The flows from horizon data beyond the wall at positive temperature end at the fold of $D(E)$ or in this collapse, and just beyond the wall at small $\phi_h$ the fold comes before the turn, with $\phi$ still close to $\phi_h$, as for the saturated flows of section \ref{sec:wall}.
The same fates occur below $\lambda^*$ outside the basin of the Lifshitz fixed point, so $\lambda^*$ removes only the Lifshitz exit.
The weak-field closed flows of Figure \ref{fig:fatemaps} also turn over at $r_s$ and end at the fold, reaching $E_{\rm sat}$ only after the turn (section \ref{sec:wall}), as the orange flow of Figure \ref{fig:blowup} shows.

The flow is therefore neither a new asymptotic geometry like the Lifshitz fixed point nor a failure at the horizon, but a spacetime that closes off before reaching any boundary (section \ref{sec:lstar}).

Figure \ref{fig:blowup} compares five flows in the proper distance $s$.
The blue flow starts from horizon data $(\kappa, E_h) = (1, 1.3)$ at $\phi_h = 0.22$ beyond the wall at $\lambda = 30$ and is strong-field closed, the orange one from the same data with $E_h$ below the wall is weak-field closed, the red one from the same data as the blue one at $\lambda = 5$, inside the basin of the Lifshitz fixed point, the purple one a saturated flow at $\lambda = 30$ close to $T = 0$ with the same $\phi_h$, and the cyan one is a source free hairy black hole at $\lambda = 30$ with $\phi_h \simeq 0.1$ and a smaller $U'(r_h)$.

Panel (a) draws the transverse size as $\pm a(s)$.
The cyan flow opens without bound towards the AdS boundary and the red one towards the Lifshitz boundary, while the blue and orange ones turn over at their $r_s$ as in (\ref{eq:amax}), close, and end at the fold, and the purple one ends at the fold before its transverse size turns.

In panel (b) $E/E_{\rm sat}$ starts above one on the blue flow and below one on the orange flow, and both reach one only at the end, while their scalars grow past the turn, the purple $E/E_{\rm sat}$ climbs to one with its scalar nearly constant, the red scalar settles at its Lifshitz value, and the cyan one decays to zero.
The Ricci scalar, not shown, tends to the AdS$_4$ value $|R| = 12$ for the cyan flow and to a constant for the red one, while for the blue flow it is finite at $r_s$ and grows by many orders of magnitude towards the end.
The cyan flow opens more slowly near the horizon because its $U'(r_h)$ is smaller.

\newpage
\bibliographystyle{JHEP}
\bibliography{LGHSC}

\end{document}